\documentclass[12pt]{article}

\usepackage{graphicx}
\usepackage{epsfig}
\begin{document}
\baselineskip=.33in

\newcommand{\be}{\begin{equation}}
\newcommand{\ee}{\end{equation}}
\newcommand{\bea}{\begin{eqnarray}}
\newcommand{\eea}{\end{eqnarray}}
\newcommand{\da}{\dagger}
\newcommand{\dg}[1]{\mbox{${#1}^{\dagger}$}}
\newcommand{\hlf}{\mbox{$1\over2$}}
\newcommand{\lfrac}[2]{\mbox{${#1}\over{#2}$}}
\newcommand{\scsz}[1]{\mbox{\scriptsize ${#1}$}}
\newcommand{\tsz}[1]{\mbox{\tiny ${#1}$}}
\newcommand{\doref}{\bf (*** REF.??? ***)}


\begin{flushright} 
gr-qc/0507052 \\
LA-UR-05-5110 \\
\end{flushright} 


\begin{flushleft}

\Large{\bf Using Early Data to Illuminate the Pioneer Anomaly}

\vspace{0.5in}

\normalsize
\bigskip 

{\bf Michael Martin Nieto${^a}$ and John D. Anderson$^b$}  \\

\normalsize
\vskip 15pt

${^a}$Theoretical Division (MS-B285), Los Alamos National Laboratory,\\
University of California,  Los Alamos, New Mexico 87545, U.S.A. \\
Email: mmn@lanl.gov \\ 
\vspace{0.25in}
$^{b}$Jet Propulsion Laboratory, California Institute of  Technology,\\
Pasadena, CA 91109, U.S.A. \\ 
Email: john.d.anderson@jpl.nasa.gov


\vspace{0.5in}
\bigskip

\end{flushleft}

\baselineskip=.33in

\begin{abstract}

Analysis of the radio tracking data from the Pioneer 10/11 spacecraft
at distances between about 20 - 70 AU from the Sun has consistently
indicated the presence of an unmodeled, small, constant, Doppler blue 
shift drift of order $6 \times 10^{-9}$ Hz/s. 
After accounting for systematics, 
this drift can be interpreted as a constant acceleration
of $a_P= (8.74 \pm 1.33) \times 10^{-8}$  cm/s$^2$   
directed {\it towards} the Sun, or perhaps as a time acceleration of 
$a_t = (2.92 \pm 0.44)\times 10^{-18}$ s/s$^2$.  
Although it is suspected that there is a systematic 
origin to this anomaly, none has been unambiguously demonstrated.  
We review the current status of the anomaly, and then point out how the
analysis of early data, which was never analyzed in detail, could allow a
more clear understanding of the origin of the anomaly, be it a
systematic or a manifestation of unsuspected physics.\\

\end{abstract}

\begin{center}
\today
\end{center}

\newpage


\section{The Pioneer missions}

Pioneer 10, launched on 3 March 1972 ET\footnote{
All times in this paper are expressed in ET, 
the ephemeris time at the spacecraft.} (2 March local time),
was the first craft launched into deep space and the first to reach an outer giant planet, Jupiter, on 4 Dec. 1973 
\cite{science}-\cite{piohist}.   The navigation to Jupiter was ground-breaking in its advances and fraught with crises, but it succeeded.

During its Earth-Jupiter cruise Pioneer 10 was still bound to the
solar system.  In this period there were two maneuvers that 
corrected rocket injection errors and put the craft on a direct trajectory 
to Jupiter. The first on 7 March 1972 produced a velocity change of 27.49 
m/s. The second on 24 March 1972 produced a much smaller velocity change of 
3.32 m/s and was the final trajectory maneuver actually required to reach Jupiter at an appropriate encounter distance.\footnote{
Even so, the trajectory was corrected further in an attempt 
to provide a radio occultation experiment for a possible detection of an 
atmosphere and ionosphere on the inner Galilean satellite Io. 
A first "Io trim" was performed on 19 September 1972, and a second on 24 
April 1973. These trim maneuvers  had to be essentially perfect if the spacecraft were to pass behind Io as viewed from Earth. Remarkably, the experiment succeeded, and an ionosphere was detected on Io at an altitude of approximately 60 to 140 km \cite{kliore1974}.}
By 9 January 1973 Pioneer 10 was at a distance of 3.40 AU (astronomical unit), beyond the asteroid belt.  Jupiter was beginning to exert a substantial gravitational influence on it.  

With Jupiter encounter, Pioneer 10 reached escape velocity from the solar system. 
Pioneer 10 passed the orbital radius of Uranus at a distance of 20.165 AU on 14 November 1979 and is headed in the general direction opposite the relative motion
of the solar system in the local interstellar dust cloud.

\begin{figure}[h!] 
    \noindent
    \begin{center}  
            \epsfig{file=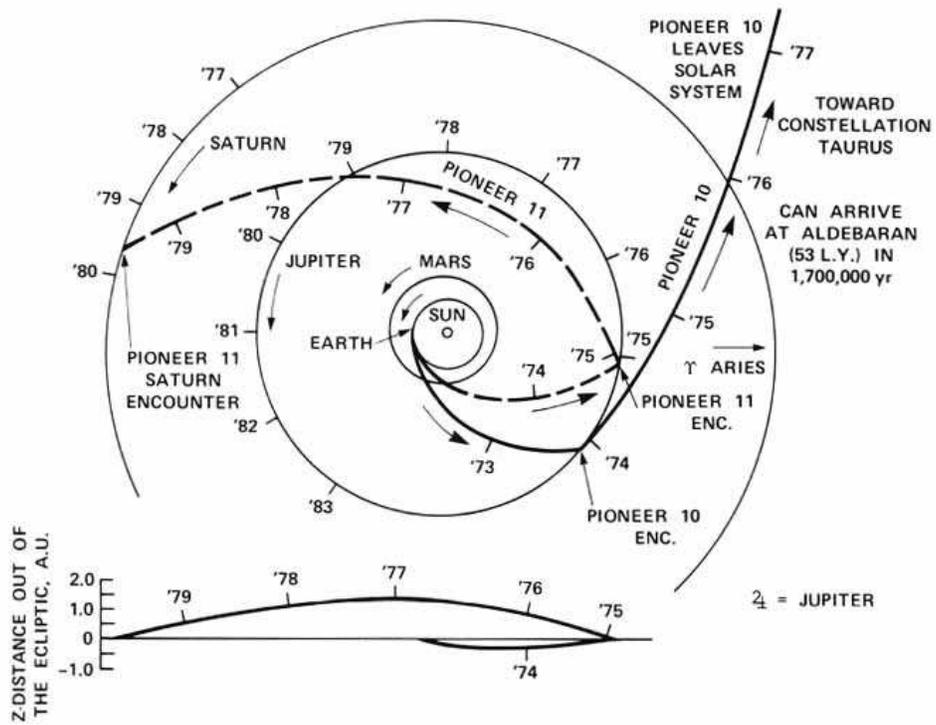,width=5in}
\caption{The Pioneer orbits in the interior of the solar system.}
 \label{earlyorbits}
    \end{center}
\end{figure}

\begin{figure}[h!] 
    \noindent
    \begin{center}  
\begin{minipage}[t]{.46\linewidth}
            \epsfig{file=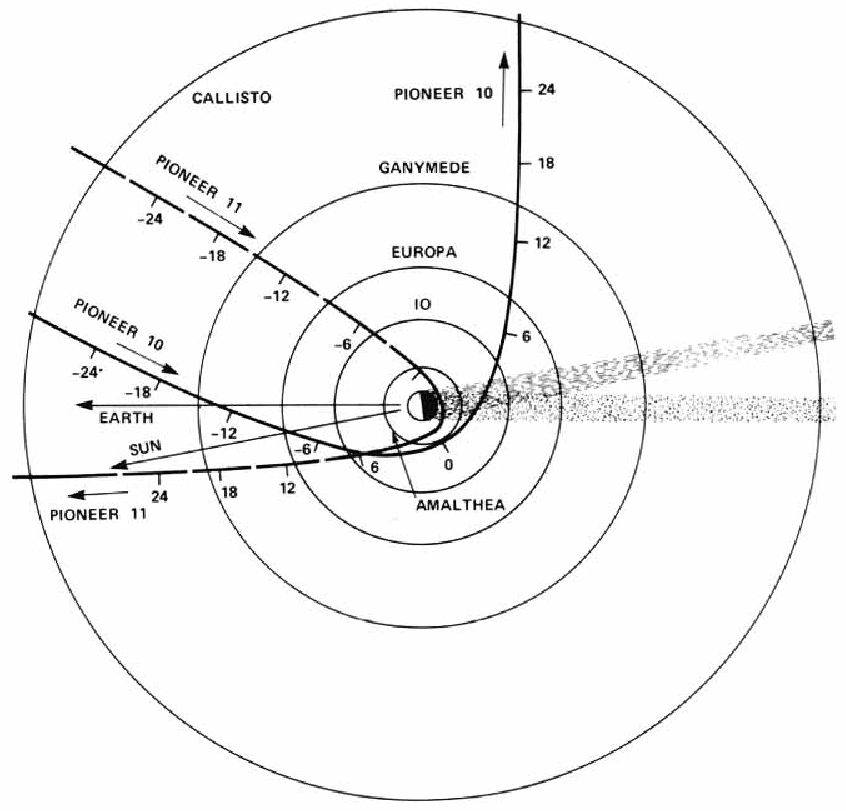,width=76mm}
            \noindent
        \end{minipage}
 \hskip 15pt
        \begin{minipage}[t]{.46\linewidth} 
            \epsfig{file=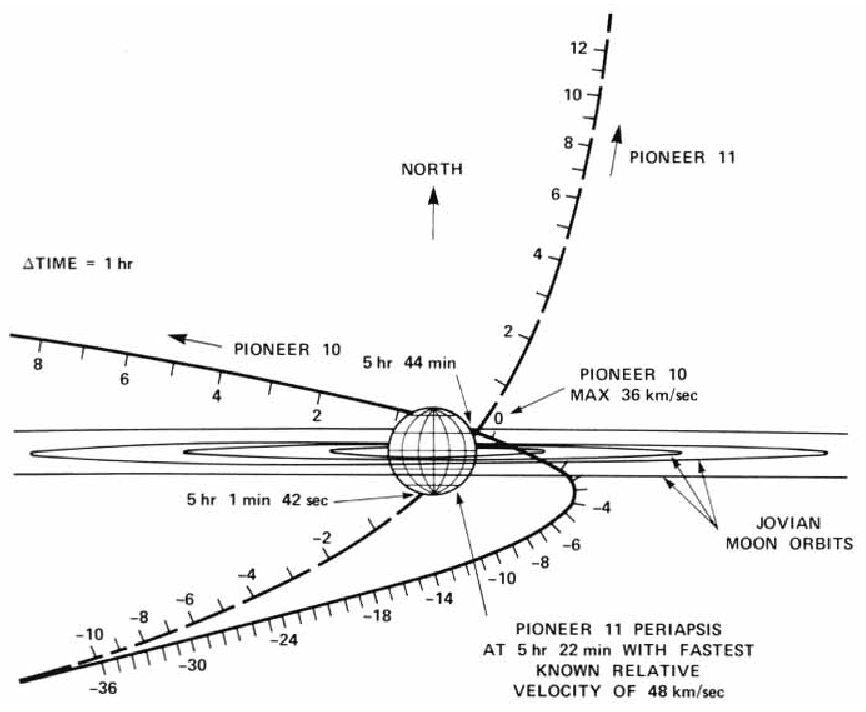,width=76mm}
            \noindent
        \end{minipage}
\caption{(a) The Jupiter flyby trajectories of the Pioneers shown from the
north pole. 
(b) The Jupiter flyby trajectories of the Pioneers shown from the
equatorial plane.}
\label{jupiterflybys}
\end{center}
\end{figure}

The Pioneer 10 solar-system orbit is shown in Figure \ref{earlyorbits}
and Figure \ref{jupiterflybys} shows the Pioneer 10 Jupiter flyby from the
polar and equatorial perspectives.
The orbital elements of Pioneer 10 for the two cruise phases are shown in 
Table \ref{pio1011cruise}.

\begin{table}[h!]
\begin{center}
\caption{\label{pio1011cruise} 
Orbital parameters for Pioneer 10 (P-10) and Pioneer 11 (P-11) during their cruises among Earth (E), Jupiter (J), and Saturn (S).  
These orbital elements are defined in standard texts in astrodynamics or celestial mechanics, for example \cite{danby88}. 
The parameters refer to the osculating orbit (ellipse or hyperbola) with respect to the solar system  barycenter at the epoch $t_0$.  
The periapse distance is $q$ and the apoapse distance is $Q$.
$v_\infty$ is the hyperbolic velocity at infinity.
$I$ is the inclination to the J2000 ecliptic.
$\Omega$ is the longitude of the ascending node.
The argument of the periapse is $\omega$ and 
$f_0$ is the true anomaly at the epoch $t_0$. 
\newline  
Except for the inclination $I$ and node $\Omega$, the orbital elements require a $GM$ for their calculation. We use the solar mass supplemented by the masses of Mercury, Earth-Moon, and Venus.  So, $GM$ for the central mass at the barycenter is $1.00000565~k^2$ (AU$^3$/day$^2$), where $k$ is Gauss' constant, an exact constant that defines the unit AU.  
}
\vskip 2pt
\begin{tabular}{|l||r|r||r|r|r|}
\hline\hline
Cruise & P-10 (E-J) & P-10 (J$\rightarrow$) &
P-11 (E-J) & P-11 (J-S) & P-11 (S$\rightarrow$)
\\[1pt]
$t_0$ [00:00 ET] &  9 Jan 1973 & 14 Nov 1979 & 
10 May 1974 & 26 Aug 1978 & 16 Dec 1985  
\\[1pt]
\hline
$q$ [AU] & $0.986$ & $5.062$ 
& $0.995$ & $3.718$ & 9.347
\\[1pt]
$Q$ [AU] & $5.851$ & $\dots$ 
& $6.053$ & $29.357$ & $\dots$
\\[1pt]
$v_\infty$ [km/s]& $\dots$ &  $11.322$ 
& $\dots$ & $\dots$ & $10.450 $ 
\\[1pt]
$v_\infty$ [AU/yr]& $\dots$ &  $2.388$ 
& $\dots$ & $\dots$ & $2.204$ 
\\[1pt]
$I$ [deg] & $2.089$ & $3.143$  
& $3.073$ & $15.320$ & $16.628$ 
\\[1pt]
$\Omega$ [deg] & $342.872$ & $332.005$  
& $16.307$ & $354.662$  & $160.353$
\\[1pt]
$\omega$ [deg] & $177.201$ & $346.769$ 
& $179.466$ & $59.976$  &  $12.798$
\\[1pt]
$f_0$ [deg] & $135.105$ & $100.457$ 
& $142.238$ & $98.987$   & $76.987$ 
\\[1pt]
\hline 
\end{tabular} 
\end{center} 
\end{table}


Pioneer 11 followed soon after with a launch on 6 April 1973 (ET), cruising 
to Jupiter on an approximate heliocentric ellipse.\footnote{
Two maneuvers were performed soon after launch.   The first, on 11 April 1973, provided a velocity change of 38.04 m/s 
and the second, on 26 April 1973, provided 1.06 m/s.}
This time during 
Earth-Jupiter cruise, it was determined that a carefully executed flyby of 
Jupiter could put the craft on a trajectory to encounter Saturn in 1979. 
Two small velocity calibrations of the cold-gas maneuver 
system were carried out on 15 March 1974 and 16 April 1974. Then, with the 
maneuver system properly calibrated, a large 65 m/s maneuver was performed 
on 19 April 1974, when Pioneer 10 was at a distance of 3.89 AU from the Sun.   This was the  maneuver that made Saturn encounter possible.  

On 2 Dec. 1974 Pioneer 11 reached Jupiter, where it underwent the Jupiter gravity assist that sent it back inside the solar system to catch up with Saturn on the far side.   It was then still on an ellipse, but a more energetic 
one.\footnote{A small calibration maneuver was 
performed a year later on 2 December 1975.  Then a 30.1 m/s trajectory 
maneuver on 18 December 1975 corrected trajectory errors 
amplified by the Jupiter flyby.} 
Pioneer 11 reached as close to the Sun as 3.73 AU on 2 February 1976.  
Another Saturn targeting maneuver of 16.6 
m/s was performed on 26 May 1976, this being the last of the large trajectory 
maneuvers for Pioneer 11.\footnote{
The final trajectory was considerably 
more energetic than required for a Saturn encounter at $\sim 10$ AU.} 
It re-crossed the orbit of Jupiter in 1977.  
After much discussion, a final small trim maneuver was performed on 
13 July 1978 for purposes of final targeting at Saturn. 
On 26 August 1978, as Saturn was beginning to substantially affect the trajectory, Pioneer 11 was at a distance of 7.509 AU. 

Pioneer 11 reached Saturn on 1 Sept. 1979.  The selected Saturn 
trajectory took the craft under the ring plane on approach and to a closest 
approach within 24,000 km of Saturn, the closest of any spacecraft to date. 
The crossing of the ring plane occurred just outside the A ring, the 
outermost of the three large visible rings with an outer edge of about 
76,500 km. 
After encounter, Pioneer 11 was on an escape hyperbolic orbit.\footnote{
Although it is imprecise to consider the total energy of the Pioneer 11 spacecraft separately from that of the rest of the solar system, if one takes the restricted 4-body problem with the solar-system barycenter as the origin of inertial coordinates and with the potential energy given by the Sun, Jupiter, and Saturn taken as point masses, Pioneer 11 reached a state of positive total energy about 2-1/2 hours before closest approach to Saturn.
}  

On 16 December 1985 Pioneer 11 was at a distance of 19.840 AU (approximately the orbital radius of Uranus).  The motion of Pioneer 11 is approximately in the direction of the Sun's relative motion in the local interstellar dust cloud (towards the heliopause).  It is roughly anti-parallel to the direction of Pioneer 10.  

\begin{figure}[t] 
\begin{center}
            \epsfig{file=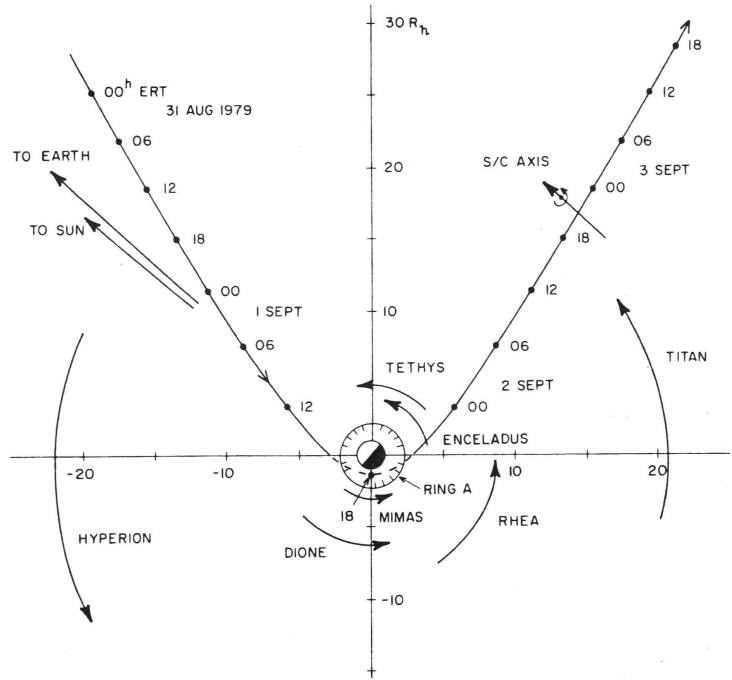,width=4in}
\caption{The Pioneer 11 Saturn flyby trajectory, projected from its 6.55$^{\circ}$ inclination on to the equatorial plane.  The curved lines show the motions of the seven major moons during the periods Pioneer 11 was interior to their orbits. 
            \label{saturnflyby}} 
    \end{center}
\end{figure}

Figure \ref{earlyorbits} shows the Pioneer 
11 interior solar system orbit.  
Figure \ref{jupiterflybys} shows the Pioneer 11 Jupiter flyby from the
polar and equatorial perspectives and Figure  \ref{saturnflyby}
shows the Pioneer 11 Saturn flyby.  
The orbital elements of Pioneer 11 for the three cruise phases are shown in 
Table \ref{pio1011cruise}.


\section{The anomaly begins to appear}
\label{appear}

Among the problems while precisely navigating in the interior of the solar system were the intense solar radiation pressure and modeling of the many gas-jet maneuvers.  (The radiation pressure caused an outward force $\sim(1/30,000)$th that of solar gravity when the craft antenna faced towards the Sun.  It varied with the spacecraft aspect.)  Even so, with measurements, calibrations, and models, both Pioneers were successfully navigated \cite{null76}.  

\begin{figure}[h]
 \begin{center}
\noindent    
\psfig{figure=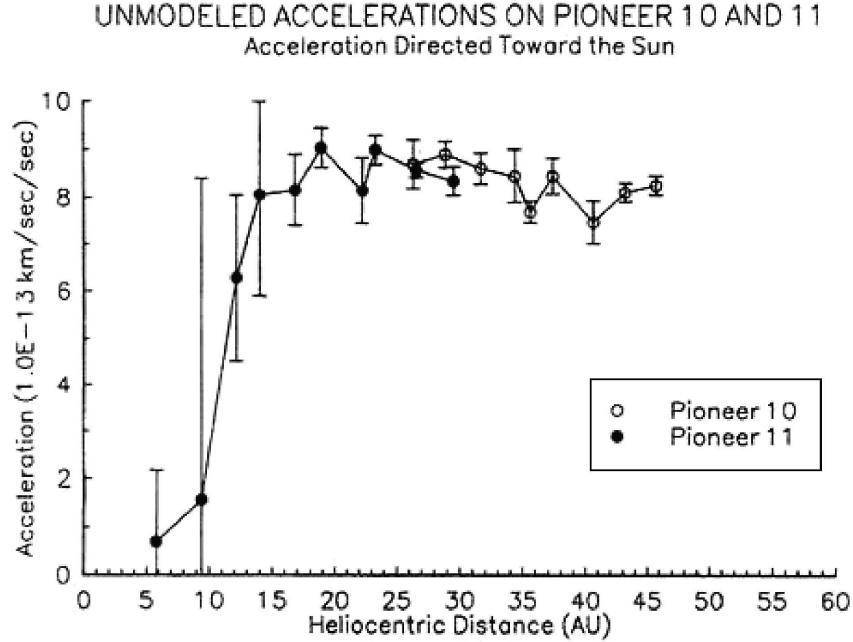,width=115mm}
\end{center}
\vskip -10pt
  \caption{A JPL Orbital Data Program (ODP) plot of the early unmodeled 
accelerations of Pioneer 10 and Pioneer 11, from about 
1981 to 1989 and 1977 to 1989, respectively.
\label{fig:correlation}}
\end{figure}

After 1976 small time-samples of data were periodically analyzed, to set limits on any unmodeled forces.  (This was especially true for Pioneer 11 which was then on its Jupiter-Saturn cruise.)  At first nothing was found.    
But when a similar analysis was done around Pioneer 11 's Saturn flyby, things dramatically changed.  (See the first two data points in Fig. 
\ref{fig:correlation}.)  So people kept following Pioneer 11.  They also started looking more closely at the incoming Pioneer 10 data.   This data taking was possible because, after their successes, an extended mission was decided upon for the Pioneers \cite{extended}.  One of the programs, of course, was continued radio-science navigation.


However, these estimates were not done in as rigorous a manner as would eventually be necessary for our study of the anomaly \cite{pioprl,pioprd}.  The raw data samples were taken by different individual analysts and put through the ``stripper/VAX" \cite{pioprd} to ``extract" the navigational data. Each individual used their own data-editing strategy, models, etc., and the points were generated from these results \cite{team}.    Further, the navigational data was not carefully archived.  That was not really necessary then because the stripper/VAX system was standard  and, in any event, the anomaly was generally believed to only be a ``curiosity." 

Even so, by 1992 an interesting string of data-points had been obtained.  They were gathered in a JPL memorandum \cite{JPLmemo} which showed a consistent bias corresponding to an acceleration of 
$\sim 8 \times 10^{-8}$ cm/s$^2$.  See Figure \ref{fig:correlation}.


\section{The modern data that was analyzed in detail}

In 1994, as part of an inquiry into how well
Newtonian gravity was known to work from intergalactic down to solar
system scales, the Pioneer data was discussed \cite{bled}.  This led to 
the long-term Pioneer Collaboration to study and understand the 
Pioneer data then in hand as well as that which was still being received
at that point \cite{pioprl,pioprd}.  

\begin{figure}[h!]
\begin{center}
\noindent    
\psfig{figure=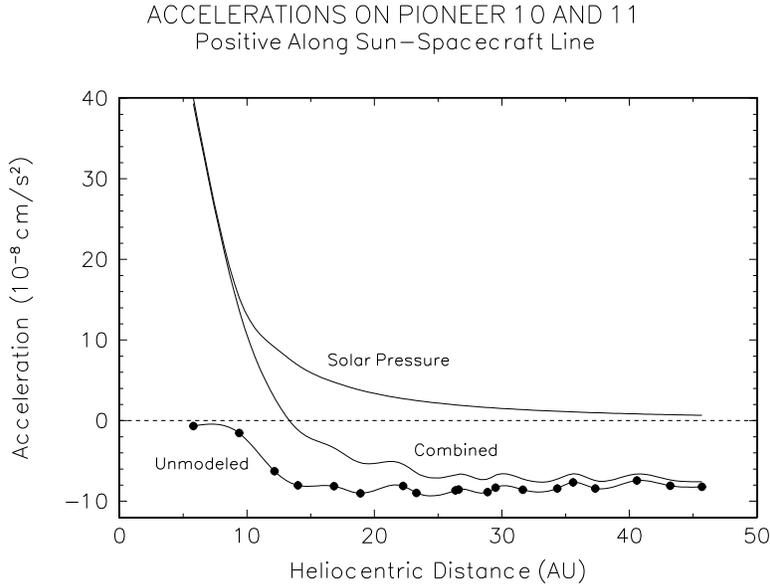,width=115mm}
\end{center}
\vskip -10pt
  \caption[ODP plots, as a function of distance from the Sun, of 
accelerations on Pioneers 10/11.]  
{ODP plots, as a function of distance from the Sun, of 
accelerations on Pioneers 10/11.  The accelerations are  a) the calculated 
solar radiation acceleration (top line), b) the unmodeled acceleration showing the data points of Figure \ref{fig:correlation} 
(bottom line), and c) the sum of the two above (middle line). 
The data points are connected by a cubic spline.
 \label{fig:forces}}
\end{figure}

The decision to seriously analyze ``modern" data,
starting from 3 January 1987 \cite{pioprl,pioprd}, 
was motivated by a number of factors.  The first was the original
concern of the community that the anomaly could well be an error in the ODP navigation code of JPL.  This concern was answered when the analysis was repeated on an independent code, The Aerospace Corporation's CHASMP navigational code \cite{pioprl}.  (Later a third code verified the anomaly \cite{mark}.)  Secondly, modern data was easily accessible in
modern format.  Finally, and perhaps most importantly, the modern data, being from far out in the
solar system, was free of many of the effects of solar system systematics 
\cite{pioprl,pioprd}, \cite{katz}-\cite{piompla}, like 
solar radiation pressure and plasma effects, which would confuse the analysis.
(See Figure \ref{fig:forces}.) 

Specifically, by 1980 Pioneer 10 was at a distance of  $\sim 20$ AU
from the Sun, the acceleration contribution from
solar-radiation pressure on Pioneer 10 (directed away from the Sun)
had decreased to $< 5 \times 10^{-8}$ cm/s$^2$.  Thus, at around this distance one could unambiguously begin to look at the data for small unmodeled forces of this magnitude.  Therefore, the fact that  
the Doppler navigational data began to indicate the presence of an anomaly 
of this order of magnitude was of more significant here. Both Pioneers had reached this distance of  $\sim 20$ AU from the Sun by 1987.


The data for Pioneer 10 was analyzed from 3 January 1987 to 22 July 1998
(40 AU to 70.5 AU). However, useful Doppler data 
continued to be received until 27 April 2002.\footnote{
It would also be useful to analyze this later Pioneer 10 data.} 
The data from Pioneer 11 was analyzed from 5 January 1987 to 1 October 1990 
(22.4 to 31.7 AU), when coherent Doppler data ceased to arrive.

The results were first reported in 1998
\cite{pioprl}. The final detailed analysis appeared in 2002 \cite{pioprd},
for which the existing Pioneer 10/11 Doppler data from 1987.0 to
1998.5 were used \cite{pioprd}. This final report extensively addressed
possible sources for a systematic origin for the detected anomaly.
The conclusion was that, even after all {\it known} systematics are
accounted for \cite{pioprd,piompla}, 
at distances between about 20 to 70 AU from the Sun 
there remains an unmodeled frequency drift of 
size\footnote{
Depending on the particular piece of data and the fitting procedure, the exact size and formal error vary slightly. (This particular number comes from the ``experimental" result for $a_P$ determined with Pioneer 10 data as described in Section VI of Ref. \cite{pioprd} and is referenced to the downlink carrier frequency, 2.29 GHz.)  But these differences are much less that the size of the anomaly and also significantly less than the systematic error.
}
$(5.99 \pm 0.01) \times 10^{-9}$ Hz/s.
This drift can be interpreted as an anomalous acceleration signal of 
$a_P=(8.74 \pm 1.33) \times 10^{-8}$  cm/s$^2$ 
in the direction towards the Sun.

We emphasize {\it known} because one might suspect that there 
is a systematic origin to the effect, perhaps generated by the
spacecraft themselves from excessive heat or propulsion gas
leaks.  But neither we nor others with spacecraft or
navigational expertise have been able to find a convincing
explanation for the anomaly \cite{pioprl,pioprd}, \cite{katz}-\cite{piompla}.


\section{Physics that the early data could illuminate}

Here we discuss the data which has and has not yet been analyzed in detail.  We emphasize how the early data, especially that taken from of order 10 AU of the Sun, may allow significantly new insights to be obtained into the origin of the anomaly. 

The original aim of the detailed analysis \cite{pioprl,pioprd} was to verify that the signal was indeed in the data,
what ever the cause, and was not an artifact of the analysis.   
The attitude was \cite{pioprd},
``Since both the gravitational and radiation
pressure forces become so large close to the Sun, the anomalous
contribution close to the Sun" (shown 
in Figures  \ref{fig:correlation} and \ref{fig:forces}) 
"is meant to represent" only 
an indication that the anomaly does exist in some form closer in to the Sun, 
but not to be a precise measurement of it.  
Many things were going on which would tend to obstruct an analysis precise to the level of the anomaly, such as many large maneuvers. 
For, example, Pioneer 11 encountered 
Jupiter and then came back across the central solar system to encounter Saturn.
The first two Pioneer 11 points were near
the distances of Jupiter and Saturn encounters.  So, effects could have been misinterpreted, for example, by imprecise modeling of maneuvers. 

Despite this, given that the anomaly is well-determined to be in the data at large distances \cite{pioprl,pioprd,mark}, we are now at a point where we have reason to reconsider this early data and to attack it critically. 

An advantage of having waited until now to investigate the noisier early data is that the analyses of the later data have given lessons on how to properly handle the data, the codes, and the external systematics, such as the solar radiation, the solar plasma,  and maneuvers.  

A difficult problem in
deep-space navigation, especially with only Doppler such as the Pioneers, 
is precise 3-dimensional orbit
determination.  The ``line-of-sight'' component of the velocity is
much more easily determined by Orbit Determination Program codes than
are the motions in the orthogonal directions. 

However, closer in to the Earth, it becomes more feasible.  A better
determination can be made of just what 
``in a direction {\it towards} the Sun'' 
exactly means.  If the anomaly is exactly: 
(i) in the solar direction this indicates a force originating from the
Sun, (ii) in the direction  towards the Earth this indicates a time 
signal anomaly,  (iii) in the ``velocity
direction" this indicates an inertial force or a drag force, or (iv) in
the spin-axis direction this indicates an on-board systematic \cite{piofind}. 
In item (iii) above we put quotes around ``velocity direction."  This is because close in to the Sun, especially for Pioneer 11 on its trans-solar-system cruise, the orbital velocity of the dust is significant with respect to that of the the craft.  Therefore, for drag this direction would be the vector sum of the spacecraft and dust velocities, and care would have to be given to determining the effective area of the craft.   

That being said, 
suppose, as expected, the Pioneer anomaly is due to internal
systematics, most often thought to be heat.\footnote{
Perhaps close in it was difficult to unmask the heat from the solar radiation pressure since there were so many maneuvers.  The effect of the heat could have been mismodeled into the maneuvers and/or the changing aspect of the craft towards the Sun.
}
Then careful analysis of the early data
should show the Pioneer anomaly being even larger than at later times,
since the radioactive decay of the RTG power sources would be less and
the electric power output from the bus would have been larger 
since the degradation of electric power was less.  (See
Ref. \cite{piompla}.)  One might even hope to tie the decrease from early
times to the two decays.  In this situation, combining {\it all} the available data (the early data, that was analyzed in detail \cite{pioprd}, and the Pioneer 10 data from 1998.5 up to last science contact in 2002) would allow the best opportunity to determine any time evolution of the anomaly.  Finally, the direction of the force
would be along the spin-axis, even when the craft was traveling almost
perpendicular to the direction towards the Earth.  3-dimensional orbit
determination might further tie this down.  

Suppose, on the other hand, that the values of the close-in  
Pioneer 11 data
points in Figures \ref{fig:forces} and \ref{fig:correlation}
represent a correct rough measurement, and not a problem with signal
to noise.  Then, especially the huge error at the second data point of
Figure \ref{fig:correlation} evaluated near Saturn flyby  could
represent a ``turn on'' of the anomaly at Saturn flyby.  

A possible new physics idea would be that near
Saturn there begins a region where there is  ``dark'' matter 
which is causing a drag force to begin. The force would then be
along the velocity vector \cite{foot,foot2}.\footnote{
What we know about ordinary dust 
indicates there is not enough to yield a large enough drag force 
further out than 20 AU \cite{drag}.  However, what effect there is would 
be expected to be larger closer in to the Sun.} 
If the ``matter" extends interior to Saturn's orbit it might 
not be as cleanly seen in the Pioneer 11 radial analysis since
it would be close to perpendicular to the Sun-spacecraft direction
during Jupiter-Saturn cruise.  Analyzing the earlier Pioneer
11 Jupiter flyby and Jupiter-Saturn cruse, as well as the Pioneer 10
Jupiter flyby would elucidate all this.  

A more radical new physics idea would be that the modified inertia
interpretation of Modified Newtonian Dynamics (MOND) is correct   
\cite{old7}.  Then the unbound hyperbolic trajectory would have an
additional acceleration of size of the Pioneer anomaly.  This would be confirmed
if the Pioneer anomaly turned on only at Saturn flyby for Pioneer 11
and only at Jupiter flyby for Pioneer 10.  
Here, then, one would want to specially analyze, 
in as good a 3-dimensional manner as possible, 
the Pioneer 10 data from about 1 year before to one year
after Jupiter flyby and the Pioneer 11 data from about one year before
Jupiter flyby to one year after Saturn flyby. 

At the least, even if the answer to the anomaly turns out to be some unknown systematic, it will still be important in the more general framework
of the solar system ephemerides and also will aid in developing protocols for spacecraft design and navigation.  


\section{Further insights from the rough early analyses}

It turns out that further insights can be gleaned from the roughly analyzed features of the early data discussed in Sec. \ref{appear}.
The data in Figure \ref{fig:correlation}  was plotted versus orbital radius in an attempt to find a common cause for the anomalous acceleration.
Although, as stated, these early acceleration points were not generated as rigorously as we might like, the ODP procedures used by all, and as extracted from numerous computer printouts and summarized in \cite{JPLmemo}, guarantee that the points are independent. They are indeed indicative of what one might find by a more rigorous analysis of the early data.\footnote{
Trajectory information good to 1\% on the Pioneers can be obtained from the 
National Space Science Data Center (NSSDC) at: \\
http://nssdc.gsfc.nasa.gov/space/helios/heli.html \\
JPL's Solar System Dynamics web site has more precise information: \\
http://ssd.jpl.nasa.gov/}  
In Table \ref{datalist} we give the information on the data points in Figure 
\ref{fig:correlation} as well as the times when the craft were at the distances quoted.  


\begin{table}[h!]
\begin{center}
\textwidth=4.25in
\caption{Pioneer 11 and 10 early data points.  Listed for the two craft are (i) the quoted values for the distance in AU, (ii) from the NSSDC the year/days-of-year at the distance position, (iii) 
the anomaly in units of $10^{-8}$ cm/s$^2$, and (iv) the error in the anomaly in the same units.  The precision of the distances as an independent variable is compatible with a precision of one day in the tabulated dates.}
\textwidth=6.25in
\label{datalist}
\begin{tabular}{|l|r|l|r|r|}   
\multicolumn{2}{c}{}\\ \hline 
Craft &       Distance & Dates ~~~~~~&   ~~~~  $a_P$  &  ~~~~  $\sigma_P$  \\  
\hline \hline 
Pioneer 11  &     5.80  &     77/270-1  &    0.69 &   1.48  \\
(Saturn Encounter)& 9.38 &    79/244    & & \\
            &     9.39  &     80/66-78  &    1.56 &   6.85  \\
            &    12.16  &     82/190-1  &    6.28 &   1.77  \\
            &    14.00  &     83/159    &    8.05 &   2.16  \\
            &    16.83  &     84/254    &    8.15 &   0.75  \\
            &    18.90  &     85/207    &    9.03 &   0.41  \\
            &    22.25  &     86/344-5  &    8.13 &   0.69  \\
            &    23.30  &     87/135-6  &    8.98 &   0.30  \\
            &    26.60  &     88/256-7  &    8.56 &   0.15  \\
            &    29.50  &     89/316-7  &    8.33 &   0.30  \\
&&&&\\
\hline
Pioneer 10  &    26.36  &     82/19     &    8.68 &   0.50  \\
            &    28.88  &     82/347-8  &    8.88 &   0.27  \\
            &    31.64  &     83/346    &    8.59 &   0.32  \\
            &    34.34  &     84/338-9  &    8.43 &   0.55  \\
            &    35.58  &     85/138    &    7.67 &   0.23  \\
            &    37.33  &     86/6-7    &    8.43 &   0.37  \\
            &    40.59  &     87/80     &    7.45 &   0.46  \\
            &    43.20  &     88/68     &    8.09 &   0.20  \\
            &    45.70  &     89/42-3   &    8.24 &   0.20  \\

\hline
\multicolumn{3}{c}{}\\[-10pt] 
\end{tabular}
\end{center}
\end{table}


Starting with the second Pioneer 11 data point, the points (which are averages of data over extended periods of many months up to approximately a year) were obtained about once a year.  In particular, the second Pioneer 11 data point  (whose stated distance of 9.38 AU was on day 1979/244, after 
Saturn encounter) comes from a data span that started before Saturn 
encounter.\footnote{  
The second Pioneer 11 data point was stated to have been taken before (or at) Saturn encounter at 9.39 AU \cite{JPLmemo}.  But since Saturn encounter was at 9.38 AU, that would mean there either was a round-off in the distance quoted or the data overlapped the encounter.  Either way the huge error in this point is anomalous and therefore it is of great interest to reanalyze this region.
} 

\begin{figure}[h]
 \begin{center}
\noindent    
\psfig{figure=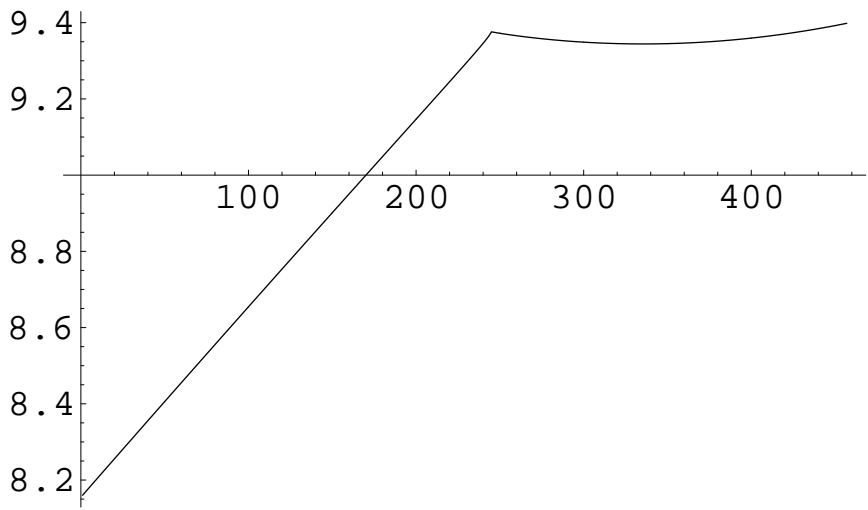,width=115mm}
\end{center}
\vskip -10pt
  \caption{A plot of Pioneer 11's distance from the Sun (in AU) vs time (in day starting with 1 Jan. 1979) near Saturn encounter (on Sept. 1, 1979).
\label{piosatenc}}
\end{figure}

After encounter, for some time Pioneer 11 traveled roughly parallel to a circular orbit with a radius close to Saturn's semi-major axis.  But its high (escape) velocity was much greater than circular velocity
so it eventually pulled away.  (See Figure \ref{piosatenc}.)


Plots of the anomaly versus time were also made from these data points.   
These showed, as might be suspected from Figures \ref{fig:correlation} and \ref{fig:forces}, a possible annual variation.  This observation would be a forerunner of the discussion in Section IX-C of \cite{pioprd}.  

Doing fits to the data points, 
the best estimate of the amplitude of the Pioneer 10 sine wave is $(0.525  \pm 0.155)\times 10^{-8}$ cm/s$^2$ and that of the Pioneer 11 wave is $(0.498 \pm 0.176)\times 10^{-8}$ cm/s$^2$ (here with the first three points omitted). 
The sine waves seem real, with, e.g.,  a 95\% probability that the Pioneer 10 amplitude lies between $0.199$ and $0.834 \times 10^{-8}$ cm/s$^2$.    
The difference in phase between the Pioneer 10 and Pioneer 11 waves is 173.2$^\circ$, similar to the angular separation of the two spacecraft in ecliptic longitude. The amplitudes are in the same proportion as the cosines of the ecliptic latitudes for the two spacecraft. 

We find it doubtful that the annual term, if confirmed, has anything to do with the Pioneer anomaly itself.  We suggest that the sine waves may be caused by a misalignment of the Pioneer orbits on the J2000 ecliptic.  
On the other hand, it is not possible to rule out the Pioneer anomaly as a cause of the annual variation. We simply do not know.  This is a further question to be addressed by the reanalysis of the early data. 

It is encouraging that the late-time end points for the early anomalous acceleration data are, after the removal of the annual term, statistically consistent with our results from a careful analysis of the later data from 1987.1 to 1998.5 \cite{pioprd}. \\


\section{Conclusion}

In summary, understanding the Pioneer anomaly, no matter what turns
out to be the answer, will be of great value scientifically.  
Even if, in the end, the anomaly is due to some
systematic, understanding this will greatly aid future mission design
and navigational protocols.  But if the anomaly is due to some
not-understood physics, the importance would be spell-binding.  The
scientific and public-interest benefits to the community that this 
program would produce could be enormous. 

Given the detailed and exhaustive work that has gone into studying the anomaly over the past decade, the above assertions are now generally understood.  The analysis of the early data would provide a relatively straight-forward and inexpensive route to obtaining critical insights into the anomaly and how (and if) one should proceed with its study. 

If, in the end, the data set contains over 25 years of Pioneer 10 data and 15 years of Pioneer 11 data, one should be able to test the hypothesis that heat is the origin of the anomaly.   In 25 years, the Pioneer RTG heat will have decreased by 18\% in an approximately linear manner.  This signal could be seen, as well as any faster fall-off coming from the electrical power decline in the main bus. 

This same data set could allow a clear determination of the annual term (not to mention the diurnal one \cite{pioprd}) by taking 50 day or shorter averages of the data over the entire period.  Assuming, as before, that the term exists, a test could then be done to see if indeed the term is caused by a misalignment of the Pioneer orbits on the J2000 ecliptic, or by something else. 

Finally, if a set of quality data is retrieved around Pioneer 11's Saturn encounter (say at least a year on either side of encounter, then one should be able to determine to at least $2 \times 10^{-8}$ cm/s$^2$ if there is a ``turn-on" at Encounter.  It will be harder to determine the direction of the anomaly but, since the craft was changing its trajectory during this period, it might be possible.  (Any useful data at the Jupiter encounters might also be useful, but they will be noisier since the positions were closer to the Sun.)

Any theories which would hope to describe the anomaly would then have clear experimental criteria to explain.


\section*{Acknowledgements}

For many helpful conversations on this topic we thank 
Hansjoerg Dittus,  Eunice L. Lau, Claus L\"ammerzahl, and Slava G. Turyshev.  
We thank Dave Lozier for updating our information on the time and magnitude of midcourse maneuvers. 
M.M.N. acknowledges support by the U.S. DOE.
The work of J.D.A. was performed at the
Jet Propulsion Laboratory, California Institute of Technology, under
contract with the  National Aeronautics and Space Administration.  



\end{document}